\documentclass[12pt]{article}
\usepackage{amscd,amsmath,amssymb,amstext,amsthm,exscale,latexsym}
\usepackage{graphicx}
\usepackage{upgreek,txfonts}
\newcommand {\al}   {\alpha}       \newcommand {\bt}  {\beta}
\newcommand {\g }   {\gamma}       
\newcommand {\dl}   {\delta}       \newcommand {\e }  {\epsilon}

\newcommand {\vf }  {\varphi}      
         
\newcommand {\Lm}   {\Lambda}      \newcommand {\Om}  {\Omega}
       
\newcommand {\pl}   {\partial}

\renewcommand {\ln}{\text{\sf\small\, ln}}

\renewcommand {\det}{\text{\sf\,\small\, det\,}}

\newcommand {\vol}  {\sqrt{{\scriptstyle |}g{\scriptstyle |}}}
\newcommand   {\ex}{\text{\sf\small\,e}}          
\newcommand   {\const}{\text{\sf\small const}}
\newcommand   {\diag}{\text{\sf\,\small diag\,}}
\newcommand {\MM}  {{\mathbb M}}   
\newcommand {\MO}  {{\mathbb O}}   
   \newcommand {\MR}  {{\mathbb R}}
\newcommand {\Bk}  {\boldsymbol{k}}

\begin{document}
\title     {Geodesics and Global Properties of the Liouville Solution in General
            Relativity with a Scalar Field}
\author    {D. E. Afanasev
            \thanks{Contact author: daniilaf@gbu.edu.cn, daniel\_afanasev@yahoo.com}\\
            \sl School of Sciences, Great Bay University, No.16 Daxue Road,\\
            Songshan Lake District, Dongguan 523000, PR China
            \\ M. O. Katanaev
            \thanks{Contact author: katanaev@mi-ras.ru}\\
            \sl Steklov mathematical institute,
            \sl ul.~Gubkina, 8, Moscow, 119991, Russia}
\date      {}
\maketitle
\begin{abstract}
One parameter family of exact solutions in General Relativity with a scalar
field has been found using the Liouville metric. The scalar field potential has exponential form. This model is interesting, because, in
particular, the solution corresponding to the naked singularity provides smooth
extension of the Friedmann universe with accelerated expansion through the zero
of the scale factor back in time. All geodesics are found explicitly. Their analysis shows
that the Liouville solutions are global ones: every geodesic is either continued
to infinite value of the canonical parameter in both directions or ends up at
the singularity at its finite value.
\end{abstract}
\section{Introduction}
Exact solutions of Einstein's equations always attract much interest of
physicists and mathematicians. Though many exact solutions are known (see e.g\
\cite{StKrMaHoHe03,GriPod09}), this field of research have still many surprises.

Recently we have found a new (Liouville) exact solution in General Relativity
minimally coupled to a scalar field with arbitrary potential
\cite{AfaKat24A}.
Two assumptions were made: (i) the metric is of Liouville type, i.e.\ it is
conformally flat, the conformal factor being the sum of arbitrary functions
depending on single coordinates, and (ii) the scalar field depends on
coordinates only through the conformal factor. The corresponding general
solution to the field equations depends on one parameter and is very simple. It
is invariant under global Lorentz transformations and, therefore, admits six
noncommuting Killing vector fields. Note, that symmetry of the space-time was not
assumed from the very beginning but arises as the consequence of the field
equations (spontaneous symmetry emergence \cite{AfaKat19,AfaKat20A}). Despite
the mathematical simplicity, the solution is of considerable interest to
physics. The obtained solutions describe,
in particular, evolution of the space-time with the naked
singularity.

Note an interesting feature of the naked singularity solution. Namely, it can be
transformed into the Friedmann form only inside the light cone with the vertex at the origin of the global coordinate system, where
it describes the accelerated expansion of the homogeneous and isotropic Universe
with constant negative curvature space sections. The scale factor is equal to
zero on the light cone, which is geodesically incomplete. Global analysis shows,
that each geodesic crosses this light cone at finite value of the canonical
parameter.  Thus, the Liouville solution provides smooth continuation of the
cosmological solution through the zero of the scale factor back in time either
to the naked singularity or to the past time infinity. We believe, that this is
the first example describing a smooth extension of the accelerated expansion
through the hyper-surface of zeroes of the scale factor.

The solution exists only for the special exponential type of the scalar field
potential. Such potentials attract much interest, though they
have some difficulties in quantization. They arise in higher-dimensional gravity
models, superstring and M-theory (see, e.g. \cite{BurBar88,Townse03}) and are
used in cosmological models \cite{Halliw87,AnCaKa11,Chakra17}.

The Liouville metric admits complete separation of variables in the
Hamilton--Jacobi equation for geodesics. It means that the geodesic equations
are Liouville integrable (have four independent envolutive conservation laws
which are quadratic in momenta). That is there are four independent Killing
tensors of second rank.

In the present paper, we integrate geodesic equations explicitly and prove that
the obtained Liouville solution is a global one: every geodesic can be either
extended to infinite value of the canonical parameter in both directions or
reaches singularity at its finite value.

For convenience of the reader, we repeat the derivation of the Liouville
solution, which is quite short, with minor extensions at the beginning of the
paper.
\section{Notation and solution}                                   \label{sgsjdk}
We consider four-dimensional space-time $\MM$ with coordinates $x=(x^\al)$,
$\al=0,1,2,3$. Let there be the Liouville metric
\begin{equation}                                                  \label{aodmkj}
  g_{\al\bt}:=\Phi(x)\eta_{\al\bt},\qquad\eta_{\al\bt}:=\diag(+1,-1,-1,-1),
\end{equation}
where the conformal factor is the sum of four arbitrary functions on single
arguments
\begin{equation}                                                  \label{ajdhtg}
  \Phi(x):=\phi_0(x^0)+\phi_1(x^1)+\phi_2(x^3)+\phi_3(x^3),
  \qquad\Phi\ne0.
\end{equation}
This is the famous Liouville metric \cite{Liouvi49} well known in mechanics,
because it admits complete separation of variables in the Hamilton--Jacobi
equations for geodesic lines even if arbitrary functions are not specified.
For some choices of arbitrary functions, metric (\ref{aodmkj}) is symmetric. For
example, if all functions are constant then we have the usual Lorentz metric up
to rescaling of coordinates. There is also such choice of functions that the
Liouville metric becomes spherically symmetric (see below). However, metric
(\ref{aodmkj}) in a general case has no Killing vectors. It means that
geodesic equations are Liouville integrable even if there is no symmetry at all.
Moreover, the variables in the wave and Dirac equation on a manifold with metric
(\ref{aodmkj}) are also completely separated (with minor corrections). Thus,
the Liouville metric is of particular interest.

The Liouville metric (\ref{aodmkj}) is conformally flat, its Weyl tensor
vanishes, and, therefore, its curvature tensor is of type {\bf 0} in Petrov's
classification \cite{Petrov69}.

Christoffel's symbols for metric (\ref{aodmkj}) are
\begin{align}                                                          \nonumber
  \Gamma_{\al\bt}{}^\g:=&\frac12\big(\pl_\al g_{\bt\dl}+\pl_\bt g_{\al\dl}
  -\pl_\dl g_{\al\bt}\big)g^{\dl\g}=
\\                                                                \label{anjxhy}
  =&\frac1{2\Phi}\big(\phi'_\al\dl_\bt^\g+\phi'_\bt\dl_\al^\g
  -\phi^{\prime\g}\eta_{\al\bt}\big).
\end{align}
Here and in what follows raising and lowering of indices are performed by
the Lorentz metric $\eta_{\al\bt}$, and the
prime denotes the derivative with respect to the corresponding arguments. The
curvature and Ricci tensors and scalar curvature have the form
\begin{align}                                                    \label{ansmahh}
  R_{\al\bt\g}{}^\dl:=&\pl_\al\Gamma_{\bt\g}{}^\dl-\pl_\bt\Gamma_{\al\g}{}^\dl
  -\Gamma_{\al\g}{}^\e\Gamma_{\bt\e}{}^\dl+\Gamma_{\bt\g}{}^\e
  \Gamma_{\al\e}{}^\dl=
\\                                                                     \nonumber
  =&\frac1{2\Phi}\big(\Phi''_{\al\g}\dl_\bt^\dl-\Phi''_{\bt\g}\dl_\al^\dl
  -\Phi''_\al{}^\dl\eta_{\bt\g}+\Phi''_\bt{}^\dl\eta_{\al\g}\big)+
\\                                                                     \nonumber
  &+\frac3{4\Phi^2}\big(-\phi'_\al\phi'_\g\dl_\bt^\dl+\phi'_\bt\phi'_\g\dl_\al^\dl
  +\phi'_\al\phi^{\prime\dl}\eta_{\bt\g}-\phi'_\bt\phi^{\prime\dl}\eta_{\al\g}
  \big)+
\\                                                                     \nonumber
  &+\frac1{4\Phi^2}\phi'_\e\phi^{\prime\e}\big(\eta_{\al\g}\dl_\bt^\dl
  -\eta_{\bt\g}\dl_\al^\dl\big),
\\                                                                \label{eassxv}
  R_{\al\g}:=&R_{\al\bt\g}{}^\bt=
  \frac1{2\Phi}\big(2\Phi''_{\al\g}+\Phi''_\e{}^\e\eta_{\al\g}\big)
  -\frac3{2\Phi^2}\phi'_\al\phi'_\g,
\\                                                                \label{egsjsd}
  R:=&g^{\al\g}R_{\al\g}=\frac3{\Phi^2}\Phi''_\al{}^\al-\frac3{2\Phi^3}\phi'_\al
  \phi^{\prime\al}.
\end{align}

If $\Phi>0$, then the signature of the metric is $(+---)$, and a scalar field
$\vf(x)$ in General Relativity with cosmological constant $\Lm$ is described by
the standard action
\begin{equation}                                                  \label{ancsgt}
  S:=\int\!dx\vol\left[R-2\Lm+\frac12g^{\al\bt}\pl_\al\vf\pl_\bt\vf-V(\vf)
  \right],\qquad g:=\det g_{\al\bt},
\end{equation}
where the potential for a scalar field $V(\vf)$ will be specified later. Now we
only assume that it is bounded from below. Variation of this action with respect to all metric components $g_{\al\bt}$ and
a scalar field $\vf(x)$ yields field equations. For the Liouville metric
(\ref{aodmkj}), (\ref{ajdhtg}), they are
\begin{align}                                                     \label{abnaas}
  R_{\al\bt}=-\frac12\pl_\al\vf\pl_\bt\vf+\frac12\Phi\eta_{\al\bt}
  (V+2\Lm),
\\                                                                \label{aqsfge}
  \eta^{\al\bt}\pl^2_{\al\bt}\vf+\frac1\Phi\phi^{\prime\al}\pl_\al\vf+\Phi V'=0,
\end{align}
which are to be solved.

We shall look for solutions of the field equations assuming that the scalar
field depends on coordinates only through the conformal factor,
$\vf(x):=\Psi\big(\Phi(x)\big)$, where $\Psi\big(\Phi\big)$ is an arbitrary
function on single argument. Then equation (\ref{abnaas}) for $\al\ne\bt$
implies
\begin{equation*}
  \frac{3\phi'_\al\phi'_\bt}{\Phi^2}=\Psi^{\prime2}\phi'_\al\phi'_\bt,
\end{equation*}
where $\Psi':=d\Psi/d\Phi$. Therefore, function $\Psi$ satisfies equation
\begin{equation*}
  \Psi^{\prime 2}=\frac3{\Phi^2}
\end{equation*}
for nonconstant functions $\phi'_\al\ne0$. Its general solution is
\begin{equation}                                                  \label{afoiuy}
  \vf:=\Psi=\pm\sqrt3\ln\big(C\Phi\big)\qquad\Leftrightarrow\qquad
  C\Phi=\ex^{\pm\vf/\sqrt3},
\end{equation}
where $C>0$ is an arbitrary integration constant. Then
Eqs.~(\ref{abnaas}), (\ref{aqsfge}) reduce to overdetermined system of equations
\begin{align}                                                     \label{ahdesj}
  \Phi''_{\al\bt}+\frac12\Phi''_\g{}^\g\eta_{\al\bt}=&\frac12\eta_{\al\bt}\Phi^2
  (V+2\Lm),
\\                                                                \label{awshgf}
  \Phi''_\g{}^\g\pm\frac{\Phi^2}{\sqrt3}V'=&0.
\end{align}
Taking the trace of Eq.~(\ref{ahdesj}), we obtain
\begin{equation}                                                  \label{ajkskh}
  3\Phi''_\g{}^\g=2\Phi^2(V+2\Lm).
\end{equation}
This equation coincides with Eq.~(\ref{awshgf}) if and only if the potential satisfies
the differential equation
\begin{equation}                                                  \label{akshfb}
  V+2\Lm\pm\frac{\sqrt3}2V'=0.
\end{equation}
It has a general solution
\begin{equation*}
  |V+2\Lm|=\ex^{\mp\frac2{\sqrt3}(\vf+\vf_0)},
\end{equation*}
where $\vf_0$ is an integration constant. Assuming that the potential is bounded
from below and this relation holds for all $\vf$ the modulus sign can be
dropped. So, the potential is
\begin{equation}                                                  \label{amdfks}
  V=-2\Lm+\ex^{\mp\frac2{\sqrt3}(\vf+\vf_0)}.
\end{equation}

Now the only equation which has to be solved is Eq.~(\ref{ahdesj}). It takes
the form
\begin{equation}                                                  \label{edkfuy}
  \Phi''_{\al\bt}=\frac16\eta_{\al\bt}\Phi^2(V+2\Lm).
\end{equation}
For the Liouville conformal factor (\ref{ajdhtg}), $\Phi''_{\al\bt}\equiv0$, when
$\al\ne\bt$. Therefore, off diagonal components of this equation are satisfied.
The diagonal components reduce to
\begin{equation} \label{phi_00ii}
\begin{split}
  \phi''_0=&~~\frac16\Phi^2(V+2\Lm),
\\
  \phi''_i=&-\frac16\Phi^2(V+2\Lm),\qquad i=1,2,3.
\end{split}
\end{equation}
Taking the sum of these equations we get
\begin{equation*}
  \phi''_0+\phi''_i=0,\qquad\forall i.
\end{equation*}
The summands depend on different arguments and hence are equal to the same
positive constant $b_0$ with opposite signs. Therefore,
\begin{equation}                                                  \label{absjlp}
\begin{split}
  \phi_0=&~~b_0\big(x^0-x^0_0\big)^2+c_0,
\\
  \phi_i=&-b_0\big(x^i-x^i_0\big)^2+c_i,
\end{split}
\end{equation}
where $x^\al_0$ and $c_\al$ are arbitrary integration constants and $b_0>0$ due to Eqs. (\ref{phi_00ii}).
Shifting the coordinates, we put $x^\al_0=0$. Moreover, rescaling the coordinates
$x^\al\mapsto kx^\al$, where $k:=b_0^{-1/4}$, constant $b_0$ can be set to
unity, $b_0=1$. Hence, the conformal factor is the Lorentz invariant quadratic
polynomial:
\begin{equation}                                                  \label{avskls}
  \Phi=s+c,
\end{equation}
where
\begin{equation*}
  s:=\eta_{\al\bt}x^\al x^\bt, \qquad c:=c_0+c_1+c_2+c_3\in\MR.
\end{equation*}
Now Eq.~(\ref{edkfuy}) reduces to
\begin{equation}                                                  \label{ajfhyt}
  2=\frac16\Phi^2(V+2\Lm),
\end{equation}
and Eq.~(\ref{awshgf}) becomes
\begin{equation}                                                  \label{adjskl}
  2=\frac1{6C^2}\ex^{\mp\frac2{\sqrt 3}\vf_0},
\end{equation}
where Eqs.~(\ref{afoiuy}) and (\ref{amdfks}) were used. It defines constant
$\vf_0$ in terms of $C$. Therefore, the potential (\ref{edkfuy}) becomes
\begin{equation}                                                  \label{ajndhs}
  V=-2\Lm+12C^2\ex^{\mp\frac2{\sqrt3}\vf}=-2\Lm+\frac{12}{\Phi^2},
\end{equation}
where the last equality is rewritten Eq.~(\ref{ajfhyt}).

To summarize, Eq.(\ref{awshgf}) is equivalent to Eq.(\ref{aqsfge}) for scalar
field (\ref{afoiuy}). Eq.(\ref{awshgf}) is the trace of Eq.(\ref{ahdesj}) for
potential (\ref{amdfks}). Finally, Eq.(\ref{ahdesj}) is solved by conformal
factor (\ref{avskls}) and constant $C$ given by Eq.(\ref{adjskl}). Therefore,
all equations are solved explicitly. Moreover, it is a general solution to the
field equations up to shifting and rescaling of coordinates, and under the
following assumptions: the scalar field and potential are nonconstant, and the
potential is bounded from below.

Thus, the solution of field equations (\ref{abnaas}) and
(\ref{aqsfge}) with ansatz $\vf:=\Psi(\Phi)$ has been found. The solution is very
simple: the conformal factor in the Liouville metric and scalar field are given by Eqs.~(\ref{avskls}) and (\ref{afoiuy}). The scalar field diverges as $s\to-c$ or $s\to\pm\infty$. It is constant on two
sheeted, $s=\const>0$, and one sheeted, $-c<s=\const<0$, hyperboloids. It is
also constant on two cones $s=0$.

So, we have obtained one parameter family of Liouville metrics
\begin{equation}                                                  \label{ansdhg}
  g_{\al\bt}=(s+c)\eta_{\al\bt},\qquad s+c\ne0,
\end{equation}
which are defined in regions $s+c>0$.
This metric is, obviously, invariant with respect to global Lorentz rotations around the origin of the coordinate system $(x^\alpha)$.
Therefore, it has six noncommuting Killing vector fields. Note, that we did not
assume any symmetry of the metric at the very beginning. It appears due to the
field equations. This phenomena was called the spontaneous symmetry emergence in
\cite{AfaKat19,AfaKat20A}. Sure, metric (\ref{ansdhg}) is also spherically
symmetric, because the rotational group is the subgroup of the Lorentz group.

The Hamilton--Jacobi equation for geodesics for metric (\ref{ansdhg}) admits
complete separation of variables and belongs to class $[0,4,0]_2$ according to
the classification given in \cite{Katana23A,Katana23B}. The geodesic Hamiltonian
equations in this coordinate system have four involutive quadratic in momenta
conservation laws in general.

It is interesting that metric (\ref{ansdhg}) nontrivially depends on coordinate
time $x^0$ and spatial coordinates, and there is no coordinate system in which
it is static even locally for $s>0$. Besides, it is not homogeneous in four
dimensions.

Curvature tensor (\ref{ansmahh}) for the obtained solution is
\begin{equation}                                                  \label{asndbv}
\begin{split}
  R_{\al\bt\g}{}^\dl=&\frac1\Phi\left(3-\frac c\Phi\right)
  \big(\eta_{\al\g}\dl_\bt^\dl-\eta_{\bt\g}\dl_\al^\dl)-
\\
  &-\frac3{\Phi^2}\big(x_\al x_\g\dl_\bt^\dl-x_\bt x_\g\dl_\bt^\dl
  -x_\al x^\dl\eta_{\bt\g}+x_\bt x^\dl\eta_{\al\g}\big).
\end{split}
\end{equation}
It tends to zero when $s\to\pm\infty$, and, therefore, the space-time is
asymptotically flat there, but we cannot say that the space-time is
asymptotically Lorentzian in these coordinates, because the metric becomes
degenerate in this limit.

The simplest curvature invariants are
\begin{align}                                                     \label{amdndh}
  R^{\al\bt\g\dl}R_{\al\bt\g\dl}=&\frac{12}{\Phi^4}\left(9-6\frac c\Phi
  +5\frac{c^2}{\Phi^2}\right),
\\                                                                \label{ajghdk}
  R^{\al\bt}R_{\al\bt}=&\frac{36}{\Phi^4}\left(3+\frac{c^2}{\Phi^2}\right),
\\                                                                \label{edgffd}
  R=&\frac{6}{\Phi^2}\left(3+\frac c\Phi\right).
\end{align}
We see, that curvature is singular only for $\Phi=0$. Besides, these invariants
tend to zero as $\Phi\to\infty$. So, the coordinates vary in regions
\begin{equation*}
  s:=\eta_{\al\bt}x^\al x^\bt>-c,\qquad\Phi>0,
\end{equation*}
When $s\to-c$ the solution becomes singular, the singularity being located on
one-sheeted $(c>0)$, two-sheeted $(c<0)$ hyperboloids, or cone $(c=0)$.

In section \ref{sqjshy}, we prove that the space-time is maximally extended along
geodesics, i.e.\ any geodesic line can be either extended to infinite value of
the canonical parameter or ends up at a singularity at its finite value.
Therefore, coordinates $x^\al$ are the global ones.

\begin{figure}[hbt]
\hfill\includegraphics[width=.45\textwidth]{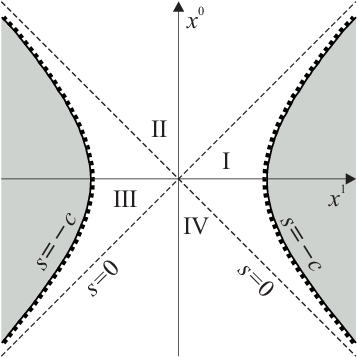}
\hfill {}
\centering\caption{The domain of definition of the Liouville metric for the
naked singularity, $\Phi>0$, $c>0$. The forbidden regions of coordinates are filled in grey.
Dashed lines correspond to ``horizons''.}
\label{Liouville}
\end{figure}

We draw the allowed regions of coordinates on the $x^0,x^1$ slice for $\Phi>0$
and $c>0$ in Fig.~\ref{Liouville}. Note, that this picture has to be
rotated (in two extra space dimensions) by the $x^0$ axis. Therefore, the
forbidden region (grey) is the connected one-sheeted ``hyperboloid''. Its
boundary is the time-like naked singularity. Test particles from the past time
infinity can either move throw the throat of the singularity and live forever or
fall on the singularity at a finite proper time. There are no space infinities.

\begin{figure}[hbt]
\hfill\includegraphics[width=.45\textwidth]{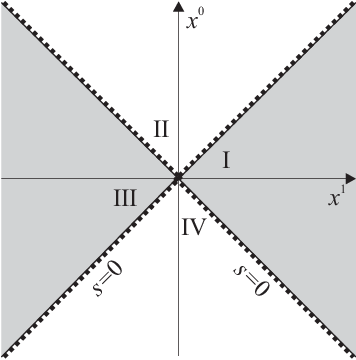}
\hfill {}
\centering\caption{The domain of coordinates definition for two cones with the
Liouville metric for $\Phi>0$, $c=0$.}
\label{Liouville0}
\end{figure}

\begin{figure}[hbt]
\hfill\includegraphics[width=.45\textwidth]{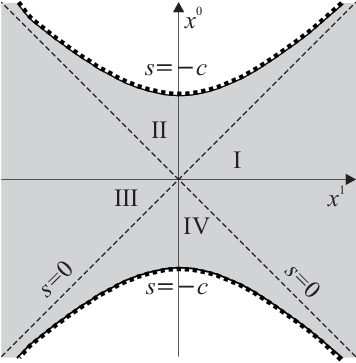}
\hfill {}
\centering\caption{The domain of coordinates definition for two hyperboloids
with the Liouville metric for $\Phi>0$, $c<0$.}
\label{Liouville-1}
\end{figure}

Figures \ref{Liouville0} and \ref{Liouville-1} show allowed regions of
coordinates in all other cases. Physical meaning of these global solutions is
not clear now.
\section{Friedman-like form of the Liouville metric}
The Liouville metric (\ref{ansdhg}) can be written in the Friedmann form by
introduction of pseudospherical coordinates in quadrant II rotated in two extra
space dimensions:
\begin{equation*}
\begin{split}
  x^0:=& t\cosh\chi,
\\
  x^1:=& t\sinh\chi\sin\theta\cos\varphiup,
\\
  x^2:=& t\sinh\chi\sin\theta\sin\varphiup,
\\
  x^3:=& t\sinh\chi\cos\theta,
\end{split}
\end{equation*}
where
\begin{equation*}
   t>0,\qquad\chi\in(0,\infty),\qquad\theta\in(0,\pi),\qquad
  \varphiup\in(0,2\pi).
\end{equation*}
The rotated quadrant IV is covered by the same coordinates with the replacement
$t\mapsto-t$.

In these coordinates, $s= t^2>0$ and the metric takes the form
\begin{equation}                                                  \label{ejslui}
  ds^2=( t^2+c)\Big(d t^2- t^2d\Om\Big),
\end{equation}
where
\begin{equation*}
  d\Om:=d\chi^2+\sinh^2\chi(d\theta^2+\sin^2\theta d\varphiup^2)
\end{equation*}
is the metric on the north sheet of two sheeted hyperboloid embedded in
Minkowskian space-time $\MR^{1,3}$ (constant negative curvature three dimensional
Riemannian manifold).

In rotated quadrants I or III, the Liouville metric cannot be written in the
Friedmann form, because hypersurfaces $s=\const$ are time-like. This means that
the Friedmann form of metric (\ref{ansdhg}) is accessible only inside the light
cone with the vertex at the origin of the coordinate system $(x^\alpha)$.

After next coordinate transformation $ t=\sqrt{\tilde t}$, $\tilde t>0$, the
Liouville metric becomes
\begin{equation}                                                  \label{ansdhf}
  ds^2=\frac14\left(1+\frac c{\tilde t}\right)d\tilde t^2-\tilde t(\tilde t+c)
  d\Om.
\end{equation}
Its limit $s\to\infty\Leftrightarrow\tilde t\to\infty$ differs from the Lorentz
metric by factor $1/4$:
\begin{equation*}
  ds^2=\frac14d\tilde t^2-\tilde t^2d\Om.
\end{equation*}
One may check that
curvature tensor components for this metric differ from zero, but its simplest
invariants (\ref{amdndh})--(\ref{edgffd}) vanish at time infinity
$\tilde t\to\infty$.

The exact metric (\ref{ejslui}) in quadrant II can be easily rewritten in the
Friedmann form too. We introduce new time coordinate $\tau(t)$ defined by the
differential equation
\begin{equation}                                                  \label{enfhjj}
  \frac{d\tau}{d t}=\sqrt{ t^2+c},\qquad t^2+c>0.
\end{equation}
Then
\begin{equation}                                                  \label{anbdgp}
  \tau-\tau_0=\frac{ t\sqrt{ t^2+c}}2+\frac c2\ln\left|
   t+\sqrt{ t^2+c}\right|,\qquad\tau_0=\text{const}.
\end{equation}
The respective metric becomes
\begin{equation}                                                  \label{ejuhgc}
  ds^2=d\tau^2-a^2(\tau)d\Om,
\end{equation}
where the scale factor
\begin{equation*}
   a(\tau)= t\sqrt{ t^2+c}
\end{equation*}
is defined implicitly by Eq.~(\ref{anbdgp}). Its derivatives are
\begin{align}                                                     \label{abansg}
  \frac{da}{d\tau}=&\frac{2 t^2+c}{ t^2+c},
\\                                                                \label{edkaan}
  \frac{d^2a}{d\tau^2}=&\frac{2c t}{( t^2+c)^{5/2}}.
\end{align}
We see, that the Universe in quadrant II is expanding with acceleration,
constant velocity and deceleration for $c>0$, $c=0$, and $c<0$, respectively.
In quadrant IV the situation is opposite.

Thus, expansion with acceleration takes place only for the naked singularity
corresponding to $\Phi>0$ and $c>0$ in quadrant II
(see Fig.~\ref{Liouville}{\em a}). This
expansion starts from the ``horizon'', $t=0$, with zero scale factor, and there
is no singularity in global coordinates. Going back in time, homogeneity and
isotropy of space sections are lost after crossing the ``horizon'', and an
observer turns out in the throat of the naked singularity. He sees signals from
the naked singularity corresponding to $s=-c$ and a hole in the center. Time-like
geodesics can go through this hole and be extended to infinite past.

Cases $c=0$ and $c<0$ for positive conformal factor are shown in
Figs.~\ref{Liouville0}{\em a} and \ref{Liouville-1}{\em a}. The
singularity at $s=-c$ may be interpreted as the Big Bang but corresponding
Universe expands with constant velocity or deceleration, respectively.

Let us compare the Liouville Universe (\ref{ejuhgc}) with the Milne
metric \cite{Milne48}. For $c=0$, Eq.~(\ref{enfhjj}) yields
\begin{equation*}
  \tau=\frac12t^2+C_0,\qquad C_0\in\MR.
\end{equation*}
Putting integration constant to zero, $C_0=0$, we get
\begin{equation}                                                  \label{egnjnf}
  ds^2=d\tau^2-4\tau^2d\Om.
\end{equation}
This expression differs from the Milne metric by factor $4$,
which is essential and cannot be removed. The Milne metric is defined on the
part of the Minkowskian space-time and has vanishing curvature. In this sense, it
is trivial: there is no matter. It describes expanding Universe with constant
velocity. Its maximal extensions along geodesics lines is the
usual flat Minkowskian space-time. In our case, maximal extension of metric (\ref{ejuhgc})
is shown in Fig.~\ref{Liouville0}{\it a}. Therefore, the Milne
metric is essentially different and not contained in expression (\ref{ansdhg}).

Similar construction can be performed in quadrants I and III. We introduce there
pseudospherical coordinates
\begin{equation}                                                  \label{abndhg}
\begin{split}
  x^0:=&r\sinh\chi,
\\
  x^1:=&r\cosh\chi\sin\theta\cos\varphiup,
\\
  x^2:=&r\sinh\chi\sin\theta\sin\varphiup,
\\
  x^3:=&r\sinh\chi\cos\theta,
\end{split}
\end{equation}
where $r>0$ is the space-like coordinate. They cover the whole one-sheeted
hyperboloid obtained by rotation of quadrant I or III in two extra space
dimensions. In these coordinates, $s=-r^2$, and the Liouville metric becomes
\begin{equation}                                                  \label{andbgf}
  ds^2=(c-r^2)(-dr^2+r^2d\Om),
\end{equation}
where
\begin{equation*}
  d\Om:=d\chi^2-\cosh^2\chi\big(d\theta^2+\sin^2\!\theta d\varphiup^2\big),
\end{equation*}
the variable $\chi$ playing the role of time. The quadratic form $d\Om$ is the
Lorentzian metric of constant curvature on the one sheeted hyperboloid. In case
$c=l^2>0$, $l>0$ (naked singularity), we introduce new space coordinate
$\rho$ by the differential equation
\begin{equation*}
  \frac{d\rho}{dr}=\sqrt{l^2-r^2},\qquad r<l.
\end{equation*}
Its general solution is
\begin{equation*}
  \rho-\rho_0=\frac{r\sqrt{l^2-r^2}}2+\frac{l^2}2\arcsin\left
  (\frac rl\right),\qquad \rho_0=\const.
\end{equation*}
Now the Liouville metric is
\begin{equation}                                                  \label{amfmfj}
  ds^2=-d\rho^2+r^2(l^2-r^2)\big[d\chi^2-\cosh^2\!\chi(d\theta^2
  +\sin^2\!\theta d\varphiup^2)\big].
\end{equation}
It is not in Friedmann-like form, because the time coordinate is $\chi$. Note
that the Liouville metric cannot be brought into the Friedmann form in quadrant
I or III, because sections $s=\text{const}$ are time-like.
\section{Geodesics for the Liouville space-time}                   \label{sqjshy}
We analyze geodesic lines for metric (\ref{ansdhg}) using the Hamilton--Jacobi
technique. The Lagrangian and Hamiltonian for time-like geodesics
$\big(x^\al(\tau)\big)$ and metric signature $(+---)$ are:
\begin{align}                                                     \label{amdnhp}
  L=&\frac12(s+c)\eta_{\al\bt}\dot x^\al\dot x^\bt,
\\                                                                \label{andjty}
  H=&\frac1{2(s+c)}\eta^{\al\bt}p_\al p_\bt,\qquad p_\al=(s+c)\eta_{\al\bt}
  \dot x^\bt,
\end{align}
where the dot denotes derivative with respect to canonical parameter
$\tau\in\MR$. Geodesic Hamiltonian equations have the form
\begin{equation}                                                  \label{abdnej}
\begin{split}
  \dot x^\al=&\frac1{s+c}p^\al,
\\
  \dot p_\al=&\frac{p_\bt p^\bt}{(s+c)^2} x_\al,
\end{split}
\end{equation}
where $p^\al:=\eta^{\al\bt}p_\bt$ and $x_\al:=\eta_{\al\bt}x^\bt$. Sure, they
are equivalent to geodesic equations in the Lagrangian formulation
\begin{equation}                                                  \label{amfnfg}
  \ddot x^\al=-\Gamma_{\bt\g}{}^\al\dot x^\bt\dot x^g
  =\frac1{s+c}(x^\al\dot x_\bt\dot x^\bt-2x_\bt\dot x^\bt\dot x^\al).
\end{equation}
These equations are clearly invariant with respect to reflection of each
coordinate $x^\al\rightarrow-x^\al$ separately.

Equations (\ref{amfnfg}) can be integrated using the Hamilton--Jacobi technique.
The Hamilton--Jacobi equation for the action function $W(x)$ is (see, i.e.\
\cite{Arnold89})
\begin{equation}                                                  \label{ehklkj}
  \frac1{s+c}\eta^{\al\bt}\pl_\al W\pl_\bt W=2m^2,\qquad s+c\ne0,
\end{equation}
where $m>0$ is the mass of a test particle moving in the space-time. The complete
integral of Eq.~(\ref{ehklkj}) depends on four independent parameters
$d=(d_\al)\in\MR^4$. The Hamilton--Jacobi equation admits complete separation of
variables, by definition, if a complete integral has the form (see, e.g.\
\cite{Katana23A})
\begin{equation*}
  W(x,d)=\sum_{\al=0}^3W_\al(x^\al,d),\qquad
  \det\frac{\pl^2W}{\pl x^\al\pl d_\bt}\ne0,
\end{equation*}
where each summand depends only on single coordinate and, possibly, a full set
of independent parameters $d$. In this case, variables in Eq.~(\ref{ehklkj}) are
completely separated as
\begin{equation}                                                  \label{eshdjm}
  W^{\prime2}_\al=2m^2x_\al^2+d_\al,\qquad\forall\al,
\end{equation}
where $W'_\al:=\pl_\al W_\al$ and $d_0-d_1-d_2-d_3=2m^2c$. We can choose, for
example, $d_1,d_2,d_3,2m^2$ as the full set of independent parameters in the
complete integral of the Hamilton--Jacobi equation (\ref{ehklkj}).

The above consideration concerns only time-like
geodesics. If the right hand side of the Hamilton--Jacobi equation
(\ref{ehklkj}) is replaced by $0$ and $-2m^2$, then it describes light-like
(null) and space-like geodesics, respectively. We can always put $2m^2=1$, $m>0$, by stretching the canonical parameter in all
cases. This means that the length of time-like and space-like geodesics is chosen
as the canonical parameter.
\subsection{Light-like (null) geodesics}                           \label{svnvkf}
Null geodesics are conformally invariant, and, therefore, they are straight lines
for the Liouville solution in coordinates $x^\al$. Nevertheless, their
completeness or incompleteness at infinities and singularities must be proved,
and this is done in the present section. To this end, we use the
Hamilton--Jacobi technique to show how it works and for comparison with the
analysis of time-like and space-like geodesics in next sections.

The Hamilton--Jacobi equation (\ref{ehklkj}) for light-like (null) geodesics,
which are the simplest ones, takes the form
\begin{equation*}
  \eta^{\al\bt}\pl_\al W\pl_\bt W=0.
\end{equation*}
Time-like geodesics must satisfy the
constraint $\dot x^\al\dot x_\al=0$, which is equivalent to $p_\al p^\al=0$ due
to Eq.~(\ref{andjty}). Therefore, the complete integral of the Hamilton--Jacobi
equation is
\begin{equation*}
  W=\sum_{\al=0}^3W_\al,
\end{equation*}
where separating functions are defined by ordinary differential equations
\begin{equation}                                                  \label{ewgjhu}
  W_\al^{\prime2}=d_\al,\qquad d_0-d_1-d_2-d_3=0.
\end{equation}
Obviously, all parameters must be positive, and therefore we denote them by
$d_\al:=k_\al^2$, $k_0>0$, $k_i\ge0$, $i=1,2,3$. Separating functions for null
geodesics are linear
\begin{equation*}
  W_\al=k_\al x^\al+\tilde b_\al,\qquad\text{(no summation)},
\end{equation*}
where $\tilde b_\al\in\MR$ are integration constants.

In the Hamiltonian formulation, there are four involutive conservation laws (all
momenta are conserved):
\begin{equation*}
  p_\al=k_\al.
\end{equation*}
They are linear but not quadratic because square root of Eq.~(\ref{ewgjhu}) can
be taken. Using definition of momenta (\ref{andjty}), conservation laws can be
rewritten as differential equations for coordinates:
\begin{equation}                                                  \label{ekjios}
  (s+c)\dot x^\al=k^\al,\qquad k^\al:=\eta^{\al\bt}k_\bt,\qquad\forall\al.
\end{equation}
Rescaling the canonical parameter, we can put $k^0=1$ without loss of
generality. Dividing these relations, we obtain equations for the form of
geodesic lines
\begin{equation*}
  \frac{dx^i}{dx^0}=k^i,\qquad i=1,2,3.
\end{equation*}
They are easily integrated
\begin{equation}                                                  \label{ejuiko}
  x^i=k^ix^0+b^i,
\end{equation}
where $b^i\in\MR$ are integration constants. We see, that all null geodesics are
straight lines going through each point of the space-time in direction of the
null vector $(k^0=1,k^i)$, i.e.\ they cross a plane $x^0=\const$ by angle
$\pi/4$. This means that null geodesics have the same form as in the Minkowskian
space-time, as expected. The difference is related to the dependence on the
canonical parameter.

To find this dependence, we have to solve, for example,
Eq.~(\ref{ekjios}) for $\al=0$. Equation (\ref{ejuiko}) implies
\begin{equation}                                                  \label{egdrwf}
  s=2kx^0+b,\qquad k:=-\dl_{ij}k^ib^j,\quad b:=-\dl_{ij}b^ib^j\le0.
\end{equation}
Then we obtain dependence on the canonical parameter
\begin{equation}                                                  \label{ehygtf}
 (2kx^0+b+c)\dot x^0=1\qquad\Leftrightarrow\qquad \tau-\tau_0=k\big(x^0\big)^2
 +(b+c)x^0,
\end{equation}
where $\tau_0$ is an integration constant, which can be put to zero. We see, that
every point with finite $x^0$ is reached by null geodesic at a finite value of
the canonical parameter. It means, in particular, that all singular points are
incomplete with respect to null geodesics. Null geodesics go to light-like
infinity only when $x^0\to\pm\infty$. The second equality (\ref{ehygtf}) implies
that $\tau\to\pm\infty$ when $x^0\to\pm\infty$ for $k\ne0$. If $k=0$ and
$b+c=0$, then $s+c=0$ due to Eq.~(\ref{egdrwf}), which is forbidden. Therefore,
light-like infinities are geodesically complete, but singularities are not.
\subsection{Time-like geodesics for $\Phi>0$}
If metric has signature $(+---)$ corresponding to positive conformal factor
$\Phi=s+c>0$, then the right hand side of the Hamilton--Jacobi equation
(\ref{ehklkj}) is equal to $2m^2$ and $-2m^2$ for time-like and space-like
geodesics, respectively. Without loss of generality, we put $2m^2=1$ by rescaling
the canonical parameter in all cases for simplicity. First, we consider time-like geodesics. Separating functions $W_\al$ in this case are defined by ordinary differential equations:
\begin{equation}                                                  \label{abdvfd}
  W_\al^{\prime2}=\big(x^\al\big)^2+d_\al>0,\qquad\forall\al,
\end{equation}
where
\begin{equation}                                                  \label{ansdfg}
  d_0-d_1-d_2-d_3=c.
\end{equation}
In this way, variables in the Hamilton--Jacobi equation (\ref{ehklkj}) are
completely separated, and hence the Liouville metric belongs to class
$[0,4,0]_2$ of separable metrics according to the classification given in
\cite{Katana23A}.

In the Hamiltonian formulation, equations for separating functions
(\ref{abdvfd}) correspond to four involutive quadratic conservation laws:
\begin{equation*}
  (p_\al)^2-\big(x^\al\big)^2=d_\al
\end{equation*}
corresponding to four Killing second rank tensors.
Sure, we have to require $\big(x^\al\big)^2+d_\al>0$ for all $\al$, which
restrict the range of coordinates for given parameters $d_\al$. To find geodesic
lines, we integrate further the respective equations for coordinates:
\begin{equation}                                                  \label{aloght}
  (s+c)^2\big(\dot x^\al\big)^2=\big(x^\al\big)^2+d_\al.
\end{equation}
It is easily checked that the velocity of all geodesics equals unity:
\begin{equation*}
  g_{\al\bt}\dot x^\al\dot x^\bt=1,
\end{equation*}
as it should be.

Equations (\ref{aloght}) have singular solutions corresponding to particles
resting at points
\begin{equation}                                                  \label{efgsgw}
  x^i=\pm\sqrt{-d_i},\qquad\forall d_i\le0.
\end{equation}
For these particles, the conformal factor must be positive:
$s+c=\big(x^0\big)^2+d_0>0$, the dependence on time being given by differential
equation following from Eq.~(\ref{aloght}):
\begin{equation*}
  \big[\big(x^0\big)^2+d_0\big]\big(\dot x^0\big)^2=1.
\end{equation*}
Choosing orientation of the canonical parameter $\dot x^0>0$, we get a general
solution
\begin{equation*}
  \tau-\tau_0=\frac{x^0\sqrt{\big(x^0\big)^2+d_0}}2+\frac{d_0}2
  \ln\left|x^0+\sqrt{\big(x^0\big)^2+d_0}\right|,
\end{equation*}
where $\tau_0$ is an integration constant and $d_0=c+d_1+d_2+d_3$. This solution
is defined for all $x^0\in\MR$ if $d_0\ge0$. For $d_0<0$, the solution exists
if $\big(x^0\big)^2>|d_0|$, where boundary points $x_0=\pm\sqrt{-d_0}$
correspond to singularity $s+c=0$.

Singular solutions (\ref{efgsgw}) satisfy original geodesic equations
(\ref{amfnfg}) for $\al=0$. However, for space indices $\al=i$ we obtain
nontrivial relations
\begin{equation*}
  \frac{x^i}{\Big(\big(x^0\big)^2+d_0\Big)^2}=0.
\end{equation*}
It means that singular solutions (\ref{efgsgw}) are not geodesics unless all
$d_i=0\Leftrightarrow x^i=0$. Only the symmetry axis $x^0$ for $d_0=c$
and $x^i=0$ is the geodesic line.

The form of general type geodesics is defined by equation
\begin{equation}                                                  \label{amsnfg}
  \left(\frac{dx^i}{dx^0}\right)^2=\frac{\big(x^i\big)^2+d_i}
  {\big(x^0\big)^2+d_0},\qquad i=1,2,3,
\end{equation}
obtained by dividing equations (\ref{aloght}). Inequality
\begin{equation*}
  \big(x^0\big)^2+d_0=s+c+\big(x^1\big)^2+d_1+\big(x^2\big)^2+d_2
  +\big(x^3\big)^2+d_3>\big(x^i\big)^2+d_i,\qquad\forall i,
\end{equation*}
implies that the right hand side of Eq.(\ref{amsnfg}) is less then unity, i.e.\
the geodesic tangent vector lies inside light cones, as it should be. Equation
(\ref{amsnfg}) is equivalent to two equations:
\begin{equation}                                                  \label{ejdhtd}
  \pm\frac{dx^i}{\sqrt{\big(x^i\big)^2+d_i}}=
  \frac{dx^0}{\sqrt{\big(x^0\big)^2+d_0}}.
\end{equation}
Their general solution is
\begin{equation}                                                  \label{afrdjg}
  \left|x^i\pm\sqrt{\big(x^i\big)^2+d_i}\right|=
  C^i\left|x^0+\sqrt{\big(x^0\big)^2+d_0}\right|,
\end{equation}
where $C^i>0$ is an integration constant and $d_\al\ne0$ for all $\al$.
Modulus signs can be dropped assuming that integration constants may have any
sign. Parameters $(d_\al)$ can be arbitrary with the only constraint
(\ref{ansdfg}), and the range of coordinates is restricted by inequalities
$\big(x^\al\big)^2+d_\al>0$.

General solution (\ref{afrdjg}) without module signs can be rewritten as
\begin{equation*}
  \sqrt{\big(x^0\big)^2+d_0}=-x^0+\frac1{C^i}
  \left(x^i\pm\sqrt{\big(x^i\big)^2+d_i}\right).
\end{equation*}
The square of this equation yields
\begin{equation}                                                  \label{andmsd}
\begin{split}
  x^0=&\frac1{2C^i}\left(x^i\pm\sqrt{\big(x^i\big)^2+d_i}\right)-\frac{C^id_0}
  {2\left(x^i\pm\sqrt{\big(x^i\big)^2+d_i}\right)}=
\\
  &=\left(\frac1{2C^i}+\frac{C^id_0}{2d_i}\right)x^i\pm
  \left(\frac1{2C^i}-\frac{C^id_0}{2d_i}\right)\sqrt{\big(x^i\big)^2+d_i}.
\end{split}
\end{equation}
In a similar way, it can be rewritten in equivalent form without $\pm$ sign
\begin{equation}                                                  \label{ammskj}
  x^i=\left(\frac{C^i}2+\frac{d_i}{2C^id_0}\right)x^0+
  \left(\frac{C^i}2-\frac{d_i}{2C^id_0}\right)\sqrt{\big(x^0\big)^2+d_0}.
\end{equation}
The $\pm$ sign in Eq.~(\ref{andmsd}) tells us that there may be two points for
given $x^i$. Equation (\ref{ammskj}) proves that every general geodesic is
infinitely smooth in their domains of definitions.

There are three degenerate cases, which should be considered separately. The
first one corresponds to $d_0=0$ and $d_i\ne0$. Then Eq.~(\ref{ejdhtd}) is
equivalent to relations
\begin{equation*}
  \pm\frac{dx^i}{\sqrt{\big(x^i\big)^2+d_i}}=\frac{dx^0}{x^0},
\end{equation*}
which has a general solution
\begin{equation}                                                  \label{andkcc}
  x^0=\frac{x^i\pm\sqrt{\big(x^i\big)^2+d_i}}{C^i},
\end{equation}
where $C^i\ne0$ is an integration constant. It can be rewritten without $\pm$
sign
\begin{equation}                                                  \label{amsbnh}
  x^i=\frac{\big(C^i x^0\big)^2-d_i}{2C^i x^0}.
\end{equation}

In the second degenerate case, $d_0\ne0$ and $d_i=0$. The general solution of
Eq.~(\ref{ejdhtd}) is
\begin{equation}                                                  \label{esaqdf}
  x^i=\frac{x^0\pm\sqrt{\big(x^0\big)^2-d_0}}{C^0},
\end{equation}
where $C^0\ne0$ is an integration constant. It is rewritten as
\begin{equation}                                                  \label{adjdgg}
  x^0=\frac{\big(C^0 x^i\big)^2-d_0}{2C^0x^i}.
\end{equation}

The third degenerate case appears, when $d_0=0$ and $d_i=0$. Then
Eq.~(\ref{ejdhtd}) has a general solution
\begin{equation}                                                  \label{amnnjh}
  x^0=C^i x^i,\qquad\text{and}\qquad x^0=\frac{C^i}{x^i},\qquad C^i\ne0,\qquad
  \text{(no summation)}.
\end{equation}
This solution corresponds to special values of parameters. For example, if
$d_0=0$ and $d_1=0$, then the remaining parameters must satisfy relation
$-d_2-d_3=1$.

\begin{figure}[hbt]
\hfill\includegraphics[width=.9\textwidth]{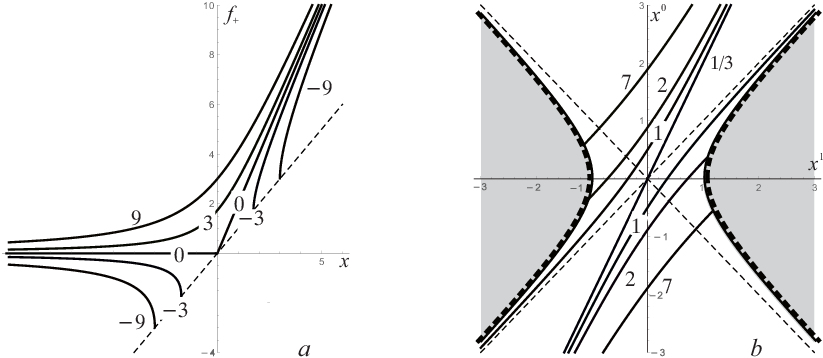}
\hfill {}
\centering\caption{Function $f_+(x)$ for different values of parameter $d$,
({\em a}). Time-like geodesics for the naked singularity, $c=1$, $C^1=1/2$, and
positive $d_1>0$ ({\em b}). The numbers denote the values of parameter $d_1$ on
both pictures.}
\label{LiouvilleGeo1}
\end{figure}

Before drawing typical time-like geodesics, we resolve modulus signs in
Eqs.~(\ref{afrdjg}) and analyse functions
\begin{equation*}
  f_\pm(x):=x\pm\sqrt{x^2+d}.
\end{equation*}
Function $f_+$ is drawn in Fig.~\ref{LiouvilleGeo1}{\it a}, for different values
of $d$. It is positive for all $d$ and $x\ge0$ and consists of two branches for
negative $d<0$ and $x\le-\sqrt{-d}$. Its asymptotics are
\begin{equation*}
\begin{aligned}
  f_+(x)\approx& ~2x+\frac d{2x}, & \quad & x\to\infty, & & & &
\\
  f_+(x)\approx& -\frac d{2x}, & & x\to-\infty, & & & &
\\
  f_+(x)\approx&~\sqrt{d} +x+\frac{x^2}{2\sqrt d}, & \quad & x\to 0, &
  \quad &d>0,
\\
  f_+(x)\approx&~\sqrt{-d}\left(1+\e+\!\sqrt{2\e}\right), & \quad &
  \e\to0_+, &\quad & d<0, & x:=&\sqrt{-d}\left(1+\e\right),
\\
  f_+(x)\approx&-\!\sqrt{-d}\left(1+\e-\!\sqrt{2\e}\right), & \quad &
  \e\to0_+, &\quad & d<0, & x:=&-\!\sqrt{-d}\left(1+\e\right).
\end{aligned}
\end{equation*}
It is clear that $f_-=-f_+(-x)$.

Let us consider geodesics of general type on the $x^1$, $x^0$ plane for
simplicity. These geodesics exist only for $d_{2,3}=0$ and $x^{2,3}=0$ due to
Eq.~(\ref{aloght}). Null geodesics on the $x^0$, $x^1$ plane are
$x^0=\pm x^1+\const$. So, light cones are the same as in the Minkowskian plane.
Equation (\ref{afrdjg}) on the $x^0,x^1$ plane is equivalent to equation
\begin{equation}                                                  \label{abdjkd}
  x^0=\left(\frac1{2C^1}+\frac{(c+d_1)C^1}{2d_1}\right)x^1\pm
  \left(\frac1{2C^1}-\frac{(c+d_1)C^1}{2d_1}\right)\sqrt{\big(x^1\big)^2+d_1}.
\end{equation}
where $d_1\ne0$ and $d_0=c+d_1\ne0$. Thus, we obtained a general solution of the
geodesic equations on the $x^0,x^1$ plain depending on two integration constants
$C^1\ne0$ and $d_1$. Note that for
\begin{equation*}
  d_1=\frac{c\big(C^1\big)^2}{1-\big(C^1\big)^2}
\end{equation*}
the general type geodesic is a straight line $x^0=x^1/C^1$.

\begin{figure}[hbt]
\hfill\includegraphics[width=.9\textwidth]{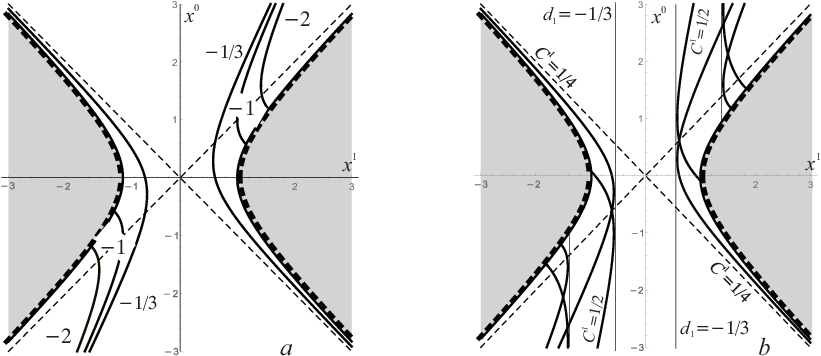}
\hfill {}
\centering\caption{Time-like general type geodesics for the naked singularity,
$c=1$, $C^1=1/2$, and different negative values of $d_1<0$ ({\em a}).
General type geodesics with the same parameter $d_1$ but different $C^1$ touch
the singular solution, which is the vertical line in Fig.~{\em b}.}
\label{LiouvilleGeo2}
\end{figure}

Time-like geodesics of general type for the naked singularity in
Fig.~\ref{Liouville}, $c=1$, $C^1=1/2$, and different positive $d_1>0$
are shown in Fig.~\ref{LiouvilleGeo1}{\it b}. For $d_1=1/3$ the geodesic is the
straight line. Sure, one has to add geodesics obtained by reflection of space
coordinate $x^1\to-x^1$ due to the $\pm$ sign in Eq.~(\ref{ejdhtd}).
Time-like geodesics for $c=1$ and negative $d_1<0$ are drawn in
Fig,~\ref{LiouvilleGeo2}. Fig.~\ref{LiouvilleGeo2}{\em a} shows general type
geodesics for $C^1=1/2$ and different values of $d_1<0$, the numbers denoting
the values of $d_1$. They are obtained by gluing two branches from
Fig.~\ref{LiouvilleGeo2}{\em a} at the points, where they touch singular solutions.
The latter are vertical lines in Fig.~\ref{LiouvilleGeo2}{\em b} enveloping general
type geodesics of the same $d_1$ but different values of $C^1$.

\begin{figure}[hbt]
\hfill\includegraphics[width=.9\textwidth]{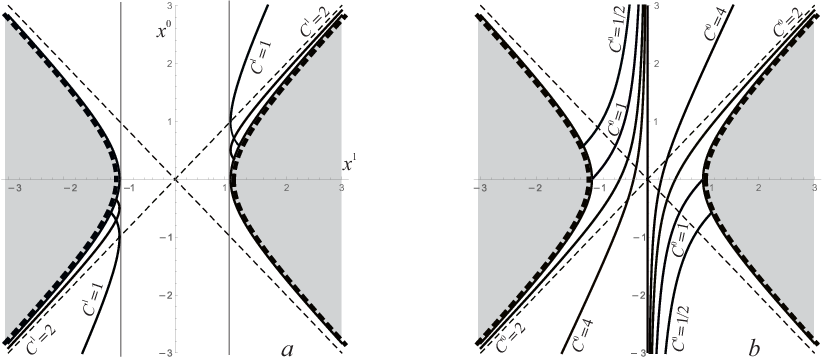}
\hfill {}
\centering\caption{Time-like degenerate geodesics of general type for the naked
singularity, $c=1$, $d_0=0$, and $d_1=-c$. They touch singular solutions
$x^1=\pm c$ for different values of $C^1$ ({\em a}). Time-like degenerate type
geodesics for $c=1$, $d_0=c$, $d_1=0$, and different values of $C^0$. They touch
the singular geodesic $x^1=0$ at infinities $x^0=\pm1$.}
\label{LiouvilleGeo3}
\end{figure}

There are also degenerate geodesics of general type. Let $d_0=0$, then $d_1=-c$,
and general solution of Eq.~(\ref{ejdhtd}) becomes
\begin{equation}                                                  \label{ahjdsy}
  x^0=\frac{x^1\pm\sqrt{\big(x^1\big)^2-c}}{C^1},
\end{equation}
where $C^1\ne0$ is an integration constant. These degenerate  geodesics smoothly
touch at points $\big(x^0=\sqrt c/C^1,x^1=\sqrt c\big)$ singular straight
solutions $x^1=\pm c$ going through vertices of hyperbolas
$\big(x^0\big)^2-\big(x^1\big)^2=c$. Equation (\ref{ahjdsy}) with $+$ and $-$
signs describes the upper and lower parts of the degenerate geodesic in
Fig.~\ref{LiouvilleGeo3}{\it a} on the right.

The second degenerate case corresponds to $d_0=c$ and $d_1=0$. Now a general
solution of Eq.~(\ref{ejdhtd}) is
\begin{equation*}
  x^1=\frac{x^0\pm\sqrt{\big(x^0\big)^2+c}}{C^0},
\end{equation*}
where $C^0\ne0$ is an integration constant. It can be rewritten as
\begin{equation}                                                  \label{asfgjy}
  x^0=\frac{\big(x^1C^0\big)^2-c}{2x^1C^0}.
\end{equation}
These geodesics are shown in Fig.~\ref{LiouvilleGeo3}{\it b} for $c=1$ and
different values of $C_0$. They touch the singular geodesic $x^1=0$ at
infinities $x^0=\pm\infty$.

There are no third degenerate time-like geodesics on the $x^0$, $x^1$ plain
because $d_0=d_1=d_2=d_3=0$ in this case, which contradicts assumption
(\ref{ansdfg}).

We see, that first integrals of geodesic equations admit singular solutions, which
are not geodesics except symmetry axis $x^0$. In spite of their existence, one
and only one time-like geodesic goes through every point of the space-time in a given
direction.

\begin{figure}[hbt]
\hfill\includegraphics[width=.9\textwidth]{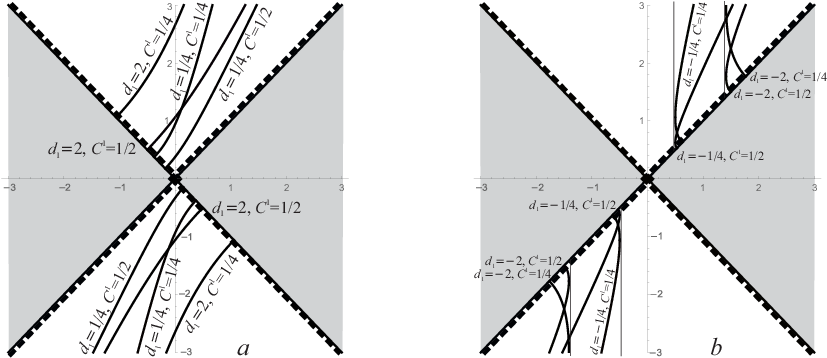}
\hfill {}
\centering\caption{Time-like general type geodesics for two cones, $\Phi>0$,
$c=0$, for different $d_1$ and $C^1$ ({\em a}). Two general type geodesics with
the same negative $d_1<0$ touch singular solutions (vertical lines) ({\it b}).}
\label{LiouvilleGeo4}
\end{figure}

Now we consider time-like geodesics for the space-time with $\Phi>0$, $c=0$
depicted in Fig.~\ref{Liouville0}.
Singular solutions are the same as in the previous case (\ref{efgsgw}) and exist
only for nonpositive $d_i\le0$.
Time-like geodesics of general type are given by Eqs.~(\ref{afrdjg}),
(\ref{andkcc}), and (\ref{esaqdf}), the only difference being the relation
between parameters $d_0-d_1-d_2-d_3=0$. Time-like geodesics of general
type on the $x^0$, $x^1$ plain are given by Eq.~(\ref{abdjkd}) with
$c=0$, which can be rewritten as
\begin{equation}                                                  \label{adejws}
  x^0=\frac12\left(\frac1{C^1}+C^1\right)x^1\pm\frac12\left(\frac1{C^1}
  -C^1\right)\sqrt{\big(x^1\big)^2+d_1},
\end{equation}
where $C^1\ne0$. Fig.~\ref{LiouvilleGeo4}{\it a} shows several geodesics of
general type for positive $d_1>0$.

General type geodesics with the same negative $d_1<0$ but different $C^1$ touch
singular vertical solutions $x^1=\pm\sqrt{-d_1}$ in
Fig.~\ref{LiouvilleGeo4}{\it b}.

\begin{figure}[hbt]
\hfill\includegraphics[width=.9\textwidth]{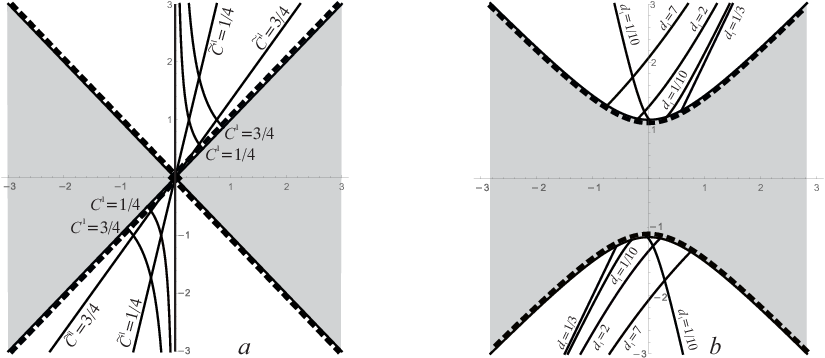}
\hfill {}
\centering\caption{Time-like degenerate geodesics of general type for two cones,
$\Phi>0$, $c=0$, for different $C^1$ and $\tilde C^1$ ({\em a}). General type
geodesics for two hyperboloids $\Phi>0$ and $c=-1$ for
$C^1=1/2$ and different positive $d_1>0$ ({\it b}).}
\label{LiouvilleGeo5}
\end{figure}

There are also degenerate geodesics of general type corresponding to
$d_0=d_1=0$:
\begin{equation*}
  x^0=\frac{x^1}{C^1},\qquad \big|C^1\big|>1,
\end{equation*}
and
\begin{equation*}
 x^0=\frac{\tilde C^1}{x^1},\qquad \big(x^1\big)^2\le \big|\tilde C^1\big|,
\end{equation*}
where inequalities follow from the requirement $\Phi>0$. They are drawn in
Fig.~\ref{LiouvilleGeo5}{\it a} for different integration constants. Degenerate
geodesics with integration constant $C^1$ are straight lines going through the
origin. The geodesics with $\tilde C^1$ are hyperbolas touching singular
geodesic $x^1=0$ at infinity.

\begin{figure}[hbt]
\hfill\includegraphics[width=.9\textwidth]{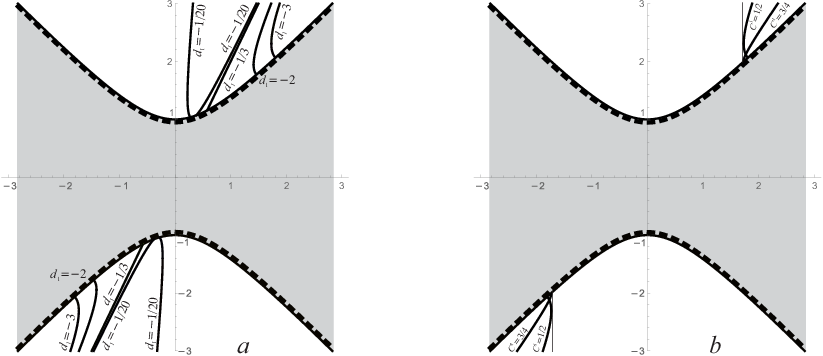}
\hfill {}
\centering\caption{Time-like geodesics of general type for two hyperboloids
$\Phi>0$ and $c=-1$ for $C^1=1/2$ and different negative $d_1<0$ ({\em a}).
General type geodesics with the same negative $d_1=-3$ and different $C^1$
touch singular solutions $x^1=\pm\sqrt3$ ({\it b}).}
\label{LiouvilleGeo6}
\end{figure}

Next we consider time-like geodesics for two hyperboloids $\Phi>0$ and $c<0$
pictured in Fig.~\ref{Liouville-1}. Geodesics of general type for
$d_0=c+d_1\ne0$, $d_1\ne0$ on the $x^0,x^1$ plain have the form (\ref{abdjkd})
as for the naked singularity. They are drawn in Fig.~\ref{LiouvilleGeo5}{\it b}
for $C^1=1/2$ and different positive values of $d_1$. Geodesics of general type
for $C^1=1/2$ and different negative values of $d_1$ are shown in
Fig.~\ref{LiouvilleGeo6}{\it a}.
General type geodesics with the same negative $d_1$ touch the singular solution
$x^1=\pm\sqrt{-d_1}$ in Fig.~\ref{LiouvilleGeo6}{\it b}.

\begin{figure}[hbt]
\hfill\includegraphics[width=.9\textwidth]{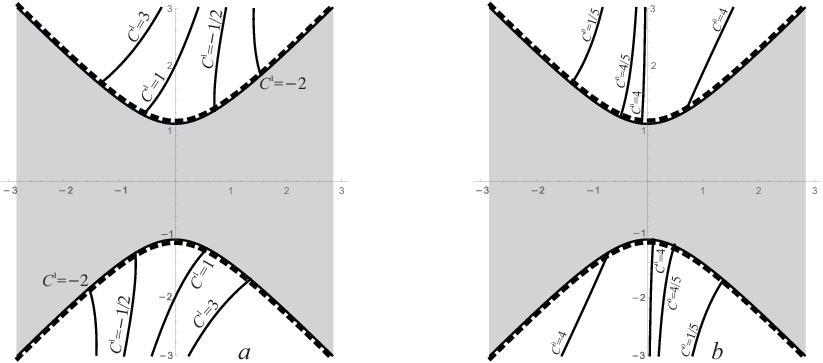}
\hfill {}
\centering\caption{Degenerate geodesics of the first kind in the space-time $\Phi>0$, $c=-1$ on the $x^0,x^1$ plain correspond to $d_0=0$ and $d_1=-c$ for different $C$ ({\em a}). Degenerate geodesics of the second kind (\ref{adjdgg}) correspond to $d_0=c$ and $d_1=0$ ({\em b}).}
\label{LiouvilleGeo7}
\end{figure}

Degenerate geodesics of the first kind in the space-time $\Phi>0$, $c=-1$ on the
$x^0$, $x^1$ plain correspond to $d_0=0$ and $d_1=-c$. They are shown in
Fig.~\ref{LiouvilleGeo7}{\it a} for different $C^1$.
Degenerate geodesics of the second kind (\ref{adjdgg}) correspond to $d_0=c$ and
$d_1=0$. They are shown in Fig.~\ref{LiouvilleGeo7}{\it b}.
\subsection{Space-like geodesics for $\Phi>0$}
The Hamilton--Jacobi equation for space-like geodesics $x^\al(\tau)$ for metric
signature $(+---)$ is
\begin{equation}                                                  \label{amnfgy}
  \eta^{\al\bt}\pl_\al W\pl_\bt W=-(s+c).
\end{equation}
The variables are completely separated by
\begin{equation}                                                  \label{eghdsa}
  W^{\prime2}_\al=-\big[\big(x^\al\big)^2+d_\al\big]\ge0,
\end{equation}
where parameters satisfy previous constraint (\ref{ansdfg}) and must be
nonpositive, $d_\al\le0$. Now coordinates vary in the finite region
$\big(x^\al\big)^2\le-d_\al$.

Equations (\ref{eghdsa}) yield first integrals for geodesic equations
\begin{equation}                                                  \label{ahjdio}
  (s+c)^2\big(\dot x^\al\big)^2=-\big(x^\al\big)^2-d_\al.
\end{equation}
They have singular space-like solution
\begin{align}                                                          \nonumber
  &x^0=\pm\sqrt{-d_0}^2,
\\                                                                \label{eshkui}
  &\big(\dl_{jk}x^j x^k+d_1+d_2+d_3\big)^2\big(\dot x^i\big)^2
  =-\big(x^i\big)^2-d_i.
\end{align}
Substitution of these expressions into geodesic equations (\ref{amfnfg}) for
$\al=0$ yields
\begin{equation*}
  0=-\frac{\sqrt{-d_0}}{(s+c)^2}.
\end{equation*}
This is a contradictory relation except $d_0=0$. It is easily verified, that
geodesic equations for space indices $\al=i$ are satisfied. So, there are
singular geodesics only for $d_0=0$, which implies $x^0=0$, the space coordinates
being defined by Eqs.~(\ref{eshkui}).

The form of space-like and time-like geodesics of general type is defined by the
same Eqs.~(\ref{amsnfg}) but different signs of $\big(x^\al\big)^2+d_\al$. These
equations do not admit degenerate space-like geodesics of general type, because
equality $d_\al=0$ implies $x^\al=0$ due to Eq.~(\ref{eghdsa}). Equations
(\ref{amsnfg}) for space-like geodesics can be rewritten in the form
\begin{equation*}
  \pm\frac{dx^i}{\sqrt{-\big(x^i\big)^2-d_i}}=
  \frac{dx^0}{\sqrt{-\big(x^0\big)^2-d_0}},\qquad d_\al\ne0.
\end{equation*}
A general solution of these equations is
\begin{equation}                                                  \label{asdfhh}
  x^0=\sqrt{\frac{d_0}{d_i}}\left(\pm x^i\cos C^i+\sqrt{-d_i-\big(x^i\big)^2}\,
  \sin C^i\right),
\end{equation}
where $C^i\in\MR$ is an integration constant.
Equation
\begin{equation*}
  -d_0-\big(x^0\big)^2=-\frac{d_0}{d_i}\left[d_i\cos^2C^i+\big(x^i\big)^2
  \cos(2C^i)\pm x^i\sqrt{-d_i-\big(x^i\big)^2}\,\sin(2C^i)\right],
\end{equation*}
is useful for verification of the solution.

There are straight space-like
geodesics if $C^i=0$ going through the origin of coordinate system
\begin{equation*}
  x^0=\pm\sqrt{\frac{d_0}{d_i}}\,x^i.
\end{equation*}
They exist only for the naked singularity and include singular geodesic for
$d_0=0$.

\begin{figure}[hbt]
\hfill\includegraphics[width=.9\textwidth]{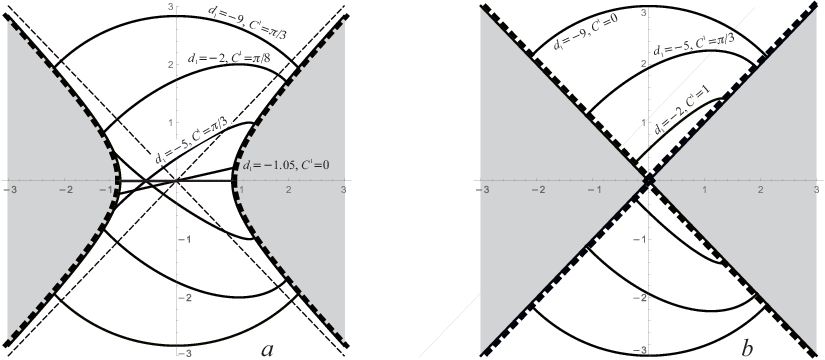}
\hfill {}
\centering\caption{Space-like geodesics of general type for the naked singularity
$c=-1$ ({\em a}) and two cones ({\it b}) for different values of $C^1$ and
$d_1<0$.}
\label{LiouvilleGeo8}
\end{figure}

\begin{figure}[hbt]
\hfill\includegraphics[width=.45\textwidth]{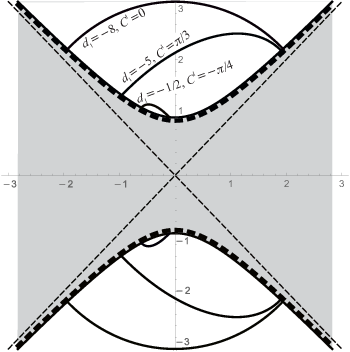}
\hfill {}
\centering\caption{Space-like geodesics of general type for two hyperboloids
for different values of $C^1$ and $d_1<0$.}
\label{LiouvilleGeo9}
\end{figure}

To visualize geodesics, we consider them on the $x^0,x^1$ plain. These geodesics
corresponding to $d_2=d_3=0$ and $d_0-d_1=c$ and zero angular momentum are
defined by equation
\begin{equation}                                                  \label{asqwdc}
  x^0=\sqrt{\frac{c+d_1}{d_1}}\left(\pm x^1\cos C^1+\sqrt{-d_1
  -\big(x^1\big)^2}\,\sin C^1\right),
\end{equation}
where coordinates are restricted by inequality $\Phi>0$. They become parts of
ellipses for $\cos C_1=0$. Figs.~\ref{LiouvilleGeo8} and \ref{LiouvilleGeo9}
show typical space-like geodesics of general type for naked singularity $(c=1)$,
two cones $(c=0)$, and two hyperboloids $(c=-1)$ for different values of $d_1$
and $C^1$. Sure, there are also reflected geodesics $x^1\to-x^1$.

\subsection{Completeness of geodesics}
Solutions in a gravity model are global ones, by definition, if every
geodesic can be either extended to infinite value of the canonical parameter or
ends up at its finite value at a singular point, where one of geometric
invariants becomes infinite. In section \ref{svnvkf}, we analysed completeness
of null geodesics, and it was proved that the obtained space-times are
maximally extended along them. But this is not enough. There are
manifolds with Lorentzian metrics (two dimensional dilaton black holes), whose
singularities are incomplete and complete for non-null and null geodesics,
respectively \cite{KaKuLi96,KaKuLi97}. Therefore, we check completeness of
non-null geodesics for the Liouville solution in the present section.

First, we consider time-like geodesics.
The equation for the canonical parameter up to its orientation follows from
Eq.~(\ref{aloght}), for example, for $\al=0$:
\begin{equation}                                                  \label{adjhki}
  d\tau=\frac{s(x^0)+c}{\sqrt{\big(x^0\big)^2+d_0}}dx^0,
\end{equation}
where we have to insert functions $x^i(x^0)$ in $s(x)$ for a given geodesic. Its
general solution is
\begin{equation}                                                  \label{aksled}
  \tau-\tau_0=\int_{x^0_0}^{x^0}\!\!\!dt\frac{s(t)+c}{\sqrt{t^2+d_0}},
\end{equation}
where the integral is taken along geodesic from point $x_0$ to $x$. For any
internal point of the space-time $\big(x^0\big)^2+d_0>0$ due to
Eq.~(\ref{abdvfd}), and, therefore, the integral converges. This means that any
internal point of the space-time is reached by geodesics at a finite value of the
canonical parameter as it should be. So, we have to analyze points at time
infinities
\begin{equation*}
  x^0\to\pm\infty,\qquad \sum_{i=1}^3\left|\frac{dx^i}{dx^0}\right|<1,
\end{equation*}
and at singular boundary.

At time infinities, time-like geodesics have asymptotics
\begin{equation*}
  x^i=k^ix^0,
\end{equation*}
where $\Bk:=(k^0=1,k^i)$ is a time-like vector. Therefore, we have
\begin{equation*}
  s(x^0)+c\simeq \left(x^0\right)^2 \Bk^2>0,\qquad x^0\to\infty,
\end{equation*}
and integral (\ref{aksled}) behaves like
\begin{equation*}
  \tau\sim\int^\infty\!\!\!dt\,t.
\end{equation*}
It is divergent, and, therefore, time infinities are geodesically complete with
respect to time-like geodesics. There are no space-like geodesics going to time
infinities.

Let time-like geodesic cross singular boundary at point $x_0=(x_0^0,x_0^i)$. Then
\begin{equation*}
  \big(x_0^0\big)^2-\big(x_0^1\big)^2-\big(x_0^2\big)^2-\big(x_0^3\big)^2
  +d_0-d_1-d_2-d_3=0,
\end{equation*}
and its asymptotic is
\begin{equation*}
  x^0=x_0^0+t,\qquad x^i=x_0^i+k^it,\qquad t\ll1,
\end{equation*}
for some time-like vector $\Bk$. Integral (\ref{aksled}) in the first
approximation takes the form
\begin{equation}                                                  \label{adhyfk}
  \tau\sim\int^0\!\!dt\,\frac{t(x_0^0-x_0^1-x_0^2-x_0^3)}
  {\sqrt{\big(x_0^0\big)^2+d_0}},
\end{equation}
because the square root differs from zero. It is convergent for nonzero
enumerator, and, therefore, time-like geodesics reach singular boundary at a finite
proper time. If $x_0^0-x_0^1-x_0^2-x_0^3=0$, then we have to retain higher order
term, but the integral remains convergent.

Space-like geodesics are incomplete at singular boundaries due to similar
arguments. Note, that all of them have a finite length. We find this property
interesting, because, in particular, the Friedmann observer sees the spatial
sections of the Universe as infinite.
Figures \ref{Liouville}, \ref{Liouville0}, and
\ref{Liouville-1} show that there are no space-like geodesics going to
space infinities. Thus, we proved that all Liouville solutions found is
the paper are global ones.
\section{Comparison with the perfect fluid stress-energy tensor}
Now we compare the energy-moment tensor of the scalar field in the naked
singularity solution with the stress-energy tensor of a perfect fluid in
quadrant II for the case $\Phi>0$ and $c>0$. Cosmological constant is put to
zero for simplicity. Then Eq.~(\ref{afoiuy}) implies
\begin{equation*}
  \pl_\al\vf=\pm\frac{2\sqrt 3 x_\al}\Phi\qquad\Rightarrow\qquad
  \pl\vf^2:=g^{\al\bt}\pl_\al\vf\pl_\bt\vf=\frac{12(\Phi-c)}{\Phi^3}.
\end{equation*}
The potential is given by Eq.~(\ref{ajndhs}). The velocity of the respective perfect fluid should be proportional to the gradient of the scalar field:
\begin{equation}
u_\alpha=\frac{\partial_\alpha\varphi}{\sqrt{\pl\vf^2}}.
\end{equation}
Thus, comparison of the scalar field with perfect fluid is relevant only
for quadrant II or IV, because the gradient of the scalar field is time-like there, while it is light-like and space-like for $s=0$ and $s<0$, respectively.  
Corresponding energy density and pressure of the scalar field are:
\begin{align}                                                     \label{energy}
  \mathcal E&=\frac12\partial\varphi^2+V=\frac{6\left(3\Phi-c\right)}{\Phi^3},
\\                                                              \label{pressure}
  \mathcal P&=\frac12\partial\varphi^2-V=-\frac{6\left(\Phi+c\right)}{\Phi^3},
\end{align}
where we assume that the square of the pseudo-radius in the argument of the
conformal factor is positive, $s>0$. Energy density and pressure are positive
and negative, respectively, and tend to zero for $s\to+\infty$. Therefore,
energy density and pressure of the scalar field vanish at time infinity. The weak
energy condition
\begin{equation*}
  \mathcal E+\mathcal P=\frac{12s}{\Phi^3}>0
\end{equation*}
holds, while the strong energy condition is violated
\begin{equation*}
  \mathcal E+3\mathcal P=-\frac{24c}{\Phi^3}<0.
\end{equation*}
According to Friedmann's equations \cite{Bauman22} the matter with such
properties provides an accelerated expansion of the Universe. Sure, this
statement is in accordance with our results, in particular, with
Eq.~(\ref{edkaan}). On the other hand, the force providing an accelerated
expansion of the Universe is associated with dark energy, which has not yet been
experimentally detected. Therefore, dark energy in the considered model is
described by the self-interacting scalar field with the exponential potential.
In contrast to cosmological constant, its energy density and pressure decrease
and vanish at the time infinity.
\section{Conclusion}
New exact one parameter family of global solutions of
Einstein equations with a scalar field is found. These solutions are
asymptotically flat at space and time infinities. The solutions are invariant
under global Lorentz transformations, nontrivially depend both on time and
space coordinates. The naked singularity solution is of particular interest. The obtained solution is spherically symmetric, because the rotational group $\MO(3)$ is the subgroup of the Lorentz group $\MO(1,3)$. As far as we know, it is essentially different from all
previously known solutions.

The Liouville metric depending on four arbitrary functions on single coordinates
is used as the initial ansatz. In general, it has no symmetry but admits four
envolutive independent indecomposable second rank Killing tensors. This is a
very important property resulting in integrability of geodesic equations.
Geodesics for obtained solutions are explicitly found and analysed. We have
proved that obtained solutions are global ones: every geodesic can be either
extended in both directions to infinite values of the canonical parameter or it
ends up at a finite value at a singular point, where curvature becomes infinite.

Analysis of time-like and space-like geodesics for conformally flat metrics in general case is far from being straightforward, because variables in the respective Hamilton-Jacobi equation are separable only for very rare metrics (all of them are listed in [13,14]). For example, if the
conformal factor depends on a function of two or more coordinates, then
coordinates are not separable in general, and a general solution of geodesic
equations cannot be given in quadratures. Moreover, if it depends on functions
of single coordinates, then geodesic equations are still not Liouville
integrable in a general situation. Only in very rare cases, it admits four
functionally independent involutive conservation laws. Besides, we are not aware
of other metrics in General Relativity with four quadratic conservation laws
which are in involution. Six linear conservation laws due to the Killing vectors
corresponding to the final Lorentz symmetry are not involutive. The advantage of
the Liouville metric is that it is separable, and we demonstrated this by direct
integration of geodesic equations. There are no pathological geodesics: all
geodesics going to infinities are complete, and geodesics at singularities are
incomplete. This proves that the obtained solution is a global one and cannot be
extended.

The naked singularity solution can be written in the Friedmann form only inside
the light cone, the zero of the scale factor being located not at a point but on
the whole light cone. Inside the cone, metric describes accelerated expansion of
the Universe, which is plausible. There is no curvature singularity on the light
cone, and the obtained global solution provides infinitely smooth continuation
of the Friedman-like metric back in time through the zero of the scale factor.
There is no Big Bang in this evolution, and the scalar field plays the role of
the dark energy. One may say, that the Liouville solution can be obtained
directly within the Friedmann initial ansatz, and this is not true. If you make
the corresponding cosmological ansatz directly into the field equations, then
you will get only local solution defined inside the light cone. It is
geodesically incomplete and requires extension. This is a separate problem, which
is far from being trivial. The Liouville ansatz leads directly to the global
solution.

Smooth continuation through the zero of the scale factor is known for a long
time \cite{Milne48}. The curvature for the Milne solution is identically zero,
and its maximal extension along geodesics results into flat Minkowskian
space-time. In this sense, it is trivial because there is no matter. The
solution obtained in our paper has nontrivial curvature, describes the naked
singularity and does not contain the Milne solution.

Most of known exact solutions in general relativity are obtained under
assumption of some symmetry of the space-time, which reduces the number of
unknown components of the metric. In the present paper, we replaced this
approach by the requirement of separability of the metric. That is the metric
must admit complete separation of variables in the Hamilton--Jacobi equation for
geodesics (the St\"ackel problem). This requirement also reduces greatly the
number of unknown metric components. Moreover, components of separable metrics
can depend only on functions of single coordinates, which means that Einstein's
equations reduce to a system of nonlinear {\em ordinary} differential equations.
Recently it was proved that there are only ten classes of separable metrics in
four dimensions \cite{Katana23A,Katana23B}, the Liouville metric belongs to one
of the classes. This approach seems to be effective, and, probably, the
Liouville solution will help us in deeper understanding of General
Relativity and, in particular, Cosmology.

D.~E. Afanasev and M.~O. Katanaev contributed equally to this work. The authors of this paper were ordered alphabetically.


\end{document}